\documentstyle[12pt]{article}

\begin{document} 
\begin{flushright}
OITS 762.1\\
May 2005
\end{flushright}
\vskip1cm

\begin{center}  {\Large {\bf Parton and Hadron Correlations in
Jets}}
\vskip .75cm
 {\bf   Rudolph C. Hwa$^1$ and  Zhiguang Tan$^{1,2}$}
\vskip.5cm

 {$^1$Institute of Theoretical Science and Department of
Physics\\ University of Oregon, Eugene, OR 97403-5203, USA\\
\bigskip
$^2$Institute of Particle Physics, Hua-Zhong Normal University,
Wuhan 430079, P.\ R.\ China}
\end{center}
\vskip.5cm

\begin{abstract} 
Correlation between shower partons is first studied in high $p_T$
jets. Then in the framework of parton recombination the correlation
between pions in heavy-ion collisions is investigated. Since thermal
partons play very different roles in central and peripheral
collisions, it is found that the correlation functions of the
produced hadrons behave very differently at different centralities,
especially at intermediate $p_T$. The correlation function that can
best exhibit the distinctive features is suggested. There is not a great deal of overlap between what we can calculate and what has been measured. Nevertheless, some aspects of our results  compare favorably with experimental data.

\vskip0.5cm
PACS numbers:  25.75.Gz, 25.75.Dw

\end{abstract}

\section{Introduction} 

Correlations between hadrons produced in jets in heavy-ion
collisions have become a subject of intense study recently
\cite{tt}-\cite{ja2}.  There are many ways to describe the nature
of the correlations, some using triggers, others giving
correlations without triggers.  So far no clear understanding
of the properties of dihadron correlations has emerged. 
Satisfactory theoretical description of the problem is lagging even farther
behind.  In this paper we offer a theoretical view of some aspect
of the correlations among partons and hadrons in connection
with jets.  Because we treat only a limited aspect of the dihadron
distributions (notably only $\pi^+\pi^+$ correlation), we cannot make exhaustive comparisons with the data that have their own limitations, such as the lack of particle identification in many cases.  Nevertheless, our results illuminate some
important features of the nature of the correlations.

For experiments at the Relativistic Heavy Ion Collider (RHIC) the
relevant range of transverse momenta ($p_T$) of the detected
hadrons in which the correlations are probed is between 0.5 and
9 GeV/c.   Since no calculation from first principles is reliable in
the intermediate $p_T$ range, any theoretical treatment of
correlations can only be done in the framework of some
phenomenological model.  The parton recombination model has
been successful in describing the single-particle inclusive
distributions in Au+Au \cite{hy} as well as d+Au collisions
\cite{hy2,hyf}.  Indeed, it has already been applied to the study
of some aspects of dihadron correlations \cite{hy3}.  We shall
extend that study and examine in more detail various features of
the correlations, not only among the produced hadrons, but also
among the partons before they hadronize.

There are limitations to what we can do here.  We shall continue
to use the one-dimensional (1D) formulation of parton
recombination, which offers a transparent display in analytical
forms the various components that contribute to dihadron
correlations, without any thing hidden in an elaborate code.  
Focusing only on the $p_T$ variable in the 1D treatment means
that we cannot consider the dependencies on the pseudorapidity
$(\eta)$ and azimuthal angle $(\phi)$ separations between the
two detected hadrons.  That shortcoming will be overcome in a
separate investigation.  But by not being distracted by the
behavior in
$\Delta \eta$ and $\Delta \phi$ variables, we can examine more
closely the separate roles that thermal and shower partons play
in the correlation problem, when only the collinear momenta are
considered.  Experimentally, they correspond to the magnitudes of
the associated particle momenta. 
Even with that simplification we have not considered all charged
particles, which are what the experiments measure without particle
identification.  To gain an overview of the general properties of 
the correlation functions, the  hadron pair $\pi^+\pi^+$ is sufficient 
and more
transparent in providing a rich display of  various features.

Our basic assumption in this work is that the shower partons are emitted independently in a jet except for the momentum constraint that the sum of the momentum fractions of all partons cannot exceed 1. This assumption is only a working hypothesis in this study and is a reasonable starting point of our investigation of correlations in order to form a base line, from which the effect of dynamical correlation can be differentiated in the future. Whatever hadronic correlation that we shall calculate in this paper therefore follows from the kinematical correlation of the shower partons in a jet. The dynamical independence of the shower partons may perhaps be approximately correct at small momentum fractions where many shower partons are likely to be present. At larger momentum fractions the kinematical constraint is important and is suitably taken into account. Since the distributions of the shower partons have been determined from the fragmentation functions (FFs) with the same assumption of dynamical independence \cite{hy4}, we are in that respect merely being self-consistent in our study of their correlation here. Since the single-particle distributions that we have calculated using those distributions of the shower partons are in good agreement with the data \cite{hy}, there is indirect evidence that our assumption may not be too far from reality.

Furthermore, another test of the reliability of the shower parton distributions (SPDs) has been carried out in  \cite{hy4.1}, where the fragmentations of $u$ and $d$ quarks to $p+\bar p$ have been calculated using the SPDs for recombination into a proton or antiproton. The resultant FFs  agree with the parametrization of KKP \cite{kkp} over 4 orders of magnitude within a factor of 2, which is roughly the margin of error made in the KKP parametrization of the data on $p+\bar p$ production \cite{hy4.1}. Because of that error in \cite{kkp}, the work on the proton FF is not included in \cite{hy4}, but it is nevertheless a demonstration that the assumption of independent shower partons except for kinematical constraint can lead to sensible result in the calculation of the proton FF, which  has not been done in any other approach to hadronization, as far as we know. In this paper we continue to use that assumption to calculate the hadronic correlations in heavy-ion collisions.

We shall first examine the single-particle distributions for Au+Au collisions 
at all centralities. The parameters determined can
then be used for subsequent studies of correlations among
partons and hadrons.  
In Sec.\ 3 the correlation among the shower partons is calculated both for
jets in vacuum and for jets produced in heavy-ion collisions
(HIC).  They behave quite differently.  Then we study in Sec.\ 4  the
correlation among hadrons, for which the role of thermal partons
becomes important in central collisions, but not as much in
peripheral collisions. In Sec.\ 5  our results are compared with existing 
experimental data in areas where such comparisons are possible.

\section{Single-Particle Distributions at all Centralities }

In \cite{hy} the single-particle distribution of hadrons produced
in Au+Au collisions at 200 GeV is studied for 0-10\% centrality
only.  We now complete the picture by considering all other
centralities, thereby determining the corresponding parameters for
the thermal partons and the value of $\xi$ that describes the
average number of hard partons emerging from the dense
medium to hadronize.  This also gives us the opportunity to
reiterate the basic equations for parton recombination.

The invariant inclusive distribution for a produced meson with
momentum $p$ is \cite{hy}
\begin{eqnarray}
p{dN_{\pi}  \over  dp} = \int {dq_1 \over  q_1}{dq_2 \over 
q_2}F_{jj^{\prime}} (q_1, q_2) R_{\pi}(q_1, q_2, p) \quad   
\label{1}
\end{eqnarray}
where the recombination function (RF) for a pion is \cite{hy5}
\begin{eqnarray}
R_{\pi}(q_1, q_2, p) ={q_1q_2  \over p^2} \delta\left({q_1 
\over p}+ {q_2  \over p} - 1\right),
\label{2}
\end{eqnarray}
The distribution for  two partons, $j$ and $j^{\prime}$, that are to
recombine has three components
\begin{eqnarray}
F_{jj^{\prime}} = {\cal TT} + {\cal TS}  +
{\cal SS} ,
\label{3}
\end{eqnarray}
where the thermal parton distribution is 
\begin{eqnarray} 
{\cal T}(q) = 
Cqe^{-q/T} ,
\label{4}
\end{eqnarray}
and the shower parton in ${\cal TS}$ has the distribution
\begin{eqnarray} 
{\cal S}(q) = \xi \sum_i \int dk k
f_i(k) S^j_i (q/k) \ ,
\label{5}
\end{eqnarray} 
whereas the two shower partons in the same jet $({\cal SS})$ have
the correlated distribution 
\begin{eqnarray} 
({\cal SS}) (q_1, q_2)=\xi\sum_i
\int dk k f_i(k) \left\{S^j_i\left({q_1\over
k}\right),S^{j'}_i\left({q_2\over k-q_1}\right)
\right\},
\label{6}
\end{eqnarray}
\begin{eqnarray}
\left\{S^j_i (x_1),\ S^{j'}_i \left({x_2\over 1-x_1}\right)\right\} =
{1\over  2} \left[S^j_i (x_1) S^{j'}_i \left({x_2\over 1-x_1}\right) +
S^j_i
\left({x_1\over 1-x_2}\right)S^{j'}_i (x_2)\right] .
\label{7}
\end{eqnarray}
The distribution $f_i(k)$ of hard parton $i$ at midrapidity in
Au+Au collisions is given in \cite{sr}.  The integral over $k$ is
from $k_{\rm min} = 3$ GeV/c to an upper limit which we shall set at
40 GeV/c.  $S^j_i$ is a matrix of shower parton distributions
(SPD)	that have been determined in \cite{hy4}.  The 
form of the two SPDs, given in Eq.\ (\ref{7}), symmetrizes the order of emission and  guarantees the
momentum conservation condition $x_1 + x_2 \leq 1$, apart
from which the shower partons are assumed to be independent. 
Thus the correlation between the partons is essentially
kinematical, as it is treated here, although other correlations of more dynamical origin can be considered later when the need arises.

The shower partons can either recombine among themselves, or
recombine with the thermal partons.  In the former case the
process reproduces the FF
\begin{eqnarray} 
xD^{\pi}_i(x) = \int {dx_1  \over  x_1} {dx_2  \over x_2}
\left\{S^j_i (x_1),\ S^{j'}_i \left({x_2 \over 1-x_1}\right)
\right\} R_{\pi} (x_1, x_2, x) \ .
\label{8}
\end{eqnarray}
Indeed, this is the equation from which the SPDs are determined
in the first place \cite{hy4}.  Thus the ${\cal SS}$ term in Eq.\ (\ref{3}),
when substituted into Eq.\ (\ref{1}), yields the fragmentation
component
\begin{eqnarray} 
p{dN_{\pi} ^{\cal SS} \over  dp} = \xi \sum_i \int dk k
f_i(k) {p
\over  k} D^{\pi}_i\left( {p  \over  k}\right),
\label{9}
\end{eqnarray}
Note that because the two shower partons are from the same jet,
Eq.\ (\ref{9}) depends linearly on $\xi f_i(k)$, which is the same
dependence that the ${\cal TS}$ component has.  Hence, when all
three terms in $F_{jj^{\prime}}$ and the RF $R_{\pi}$ are
substituted into Eq.\ (\ref{1}), we have, in the notation $p =
p_T$, 
\begin{eqnarray}
{dN_{\pi}  \over  pdp}= {C^2  \over  6}e^{-p/T} + {\xi  \over p}
\sum_{ij} \int dk k f_i(k)\left[{1  \over  k}D^{\pi}_i \left({p 
\over  k}\right) + {C  \over  p^2} \int dq_1 S^j_i\left({q_1 
\over  k}\right)(p-q_1)e^{-(p-q_1)/T}\right].
\label{10}
\end{eqnarray}
This is the basic formula for the inclusive distribution of pion. 
The hard parton $i$ is summed over all relevant species, while $j$
is summed over the two possible constituent quark species of the
pion.  The quark $j$ picks up a partner from the thermal
environment to form the pion.  That ${\cal TS}$ component
turns out to be dominant in the intermediate $p_T$ region ($3 <
p_T < 9$ GeV/c) in central Au+Au collisions, and is responsible
for the high $p/\pi$ ratio, when the production of proton is
considered in a similar way \cite{hy}.

We now apply Eq.\ (\ref{10}) to non-central Au+Au collisions.  The
hard parton distribution $f_i(k)$ is scaled down from central collisions
by the average number of binary collisions $\left< N_{\rm
coll}\right>$.  Our procedure is to fit the low-$p_T$ data (for
$p_T< 2$ GeV/c) by the first term that arises from ${\cal TT}$
recombination, thereby determining $C$ and $T$ for each centrality. 
The only remaining parameter, $\xi$, is then determined by the overall
normalization of the spectrum at each centrality so that the
shape of the large $p_T$ dependence is a prediction of Eq.\
(\ref{10}).  The data used are the $\pi^0$ spectra from
PHENIX that extend up to  $p_T = 10$ GeV/c \cite{ph}.  We, however,
calculate the $\pi^+$ distribution with $j= u$ and $\bar{d}$,
since our formalism does not distinguish the two.  Figure 1 shows
the result of our calculation for four representative centralities,
although we have treated all nine centralities, for which the
parameters are shown in Fig.\ 2.  Evidently, the spectra are well
fitted at all centralities.  The inverse slope $T$ shows no
significant dependence on centrality; the common value may be
set at 
\begin{eqnarray}
T = 0.305  \, {\rm GeV/c}.
\label{11}
\end{eqnarray} 
The value of $\xi$ increases from 0.09 at 0-10\%
 to 0.97 at 80-92\% centrality.  Thus Au+Au collision at 80-92\% is
essentially equivalent to $pp$ collision, since the suppression
factor due to energy loss is nearly 1, i.e., no jet quenching.  The
value of $C$ decreases from 21.8 (GeV/c)$^{-1}$ at 0-10\% to
2.3 (GeV/c)$^{-1}$ at 80-92\%.  The thermal component in
peripheral collision is therefore much reduced, as expected.

The contributions from the various components are shown in
Fig.\ 3 for 0-10\% and 80-92\% centralities.   It is clear that at
0-10\% the component from thermal-shower recombination is
dominant in the $3 < p_T < 9$ GeV/c range, but at 80-92\% that
component is lower than both thermal-thermal and
shower-shower components.  Indeed, ${\cal SS}$ dominates for
all $p_T > 3$ GeV/c; it means that the fragmentation model is
quite adequate for $pp$ collisions in that $p_T$ range.

\section{Correlation of Partons in Jets}

Before we consider the correlation between hadrons produced
in jets, let us first study the simpler problem of correlation
between partons in jets.  Although the latter is not measurable, it
provides a striking view of the difference between jets produced
in vacuum, as in $e^+e^-$ annihilation, and jets produced in HIC
where the hard parton momentum is not fixed.

To define our notation, let $\rho_1 (1)$ and $\rho_2 (1,2)$ be
the single-particle and two-particle distributions, and we define
the two-particle correlation function as 
\begin{eqnarray}
C_2 (1,2) = \rho_2 (1,2) - \rho_1 (1)\rho_1 (2) \quad .
\label{12}
\end{eqnarray}
Later, when we apply this formula to hadron correlation, the experimentally measurable quantities are
\begin{eqnarray}
\rho_1(1)={dN_{\pi_1}\over p_1dp_1} ,   \hskip1cm
\rho_2(1,2)={dN_{\pi_1\pi_2} \over p_1dp_1p_2dp_2} .   \label{12a}
\end{eqnarray}
The normalized correlation function is
\begin{eqnarray}
K_2 (1,2) = {C_2 (1,2) \over  \rho_1 (1)\rho_1 (2)} = r_2 (1,2) -
1 \quad ,
\label{13}
\end{eqnarray}
where 
\begin{eqnarray}
r_2 (1,2) = {\rho_2 (1,2) \over  \rho_1 (1)\rho_1 (2)}  \quad .
\label{14}
\end{eqnarray}
For reasons that will become evident below, it is 
useful to consider the ratio
\begin{eqnarray}
G_2 (1,2) = {C_2 (1,2) \over  \left[\rho_1 (1)\rho_1
(2)\right]^{1/2}}  \quad ,
\label{15}
\end{eqnarray}
a quantity that is  similar to the Pearson's correlation \cite{we}
under certain assumptions, and is closely related to a quantity considered in \cite{jp}

\subsection{Two Shower Partons in a Jet in Vacuum}

When a hard parton $i$ with any fixed  large momentum $k$
fragments in vacuum to a shower of semi-hard and soft partons,
we denote the single- and two-parton distributions in terms of the
momentum fractions $x_1$ and $x_2$ by
\begin{eqnarray}
\rho_1 (1) = S^j_i (x_1)
  \quad ,
\label{16}
\end{eqnarray}
\begin{eqnarray}
\rho_2 (1,2)=  (SS)^{jj^{\prime}}_i (x_1,x_2)
  \quad ,
\label{17}
\end{eqnarray}
where the RHS of Eq.\ (\ref{17}) is to be identified with Eq.\
(\ref{7}).  For definiteness we consider $j = u$ and $j^{\prime} =
\bar{d}$.  For $i = g$ we get for $r_2(1, 2)$ a 2D distribution as
shown in Fig.\ 4.  Other distributions with $i = u$ or $\bar{d}$
are very similar.

Since $r_2(1, 2) = 1$ implies no correlation, we see from Fig.\ 4
that the two shower partons are uncorrelated when $x_1
\approx x_2 \approx 0$, but are highly correlated at higher
$x_i$.  Kinematical constraint due to momentum conservation
forces $r_2(1, 2)$ to vanish at $x_1 + x_2 =1$.  It approaches
zero faster, as $x_1 = x_2 \to 0.5$, than it does when $x_2$ is
fixed and small, as $x_1 \to 1 - x_2$.  The implication is that
if these two partons are to recombine to form a pion at  $x =x_1 +
x_2$, there is more contribution from the asymmetric pairing of
$x_1$ and $x_2$ than from the unnecessarily stringent
requirement that $x_1 = x_2= x/2$.

\subsection{Two Shower Partons in a Jet in HIC}

For each event in HIC with a high $p_T$ jet, we can calculate the
two-parton distribution as above.  However, the hard parton
momentum $k$ cannot be fixed in an experiment.  The
distribution in the shower parton momenta $q_1$ and $q_2$,
when averaged over all events, necessitates an integration over
$k$ and a sum over all $i$.  The single-parton distribution
\begin{eqnarray}
\rho_1 (1) =  {\cal S}^j (q_1)
\label{18}
\end{eqnarray}
is given by Eq.\ (\ref{5}), while the two-parton distribution 
\begin{eqnarray}
\rho_2 (1,2)=  ({\cal SS})^{jj^{\prime}} (q_1,q_2)
\label{19}
\end{eqnarray}
is given by Eqs.\ (\ref{6}) and (\ref{7}).  Some details in the
calculation need to be explained.  When $i$ in Eq.\ (\ref{6}) is
$g$, then $S^j_iS^{j^{\prime}}_i = GG$, but when $i = u$, say,
then $S^j_iS^{j^{\prime}}_i = KL$, where $G$, $K$ and $L$ are
SPDs initiated by gluon, valence quark and sea quark,
respectively, in the notation of \cite{hy4}.  It turns out that the
gross features of the results do not depend sensitively on these
details.

For Au+Au collisions at 200 GeV we have calculated $r_2(1,2)$. 
The results for most central and most peripheral collisions are
shown in Fig.\ 5.  The main feature is that $r_2(1,2)$ is large
and increases with $q_1$ and $q_2$.  That is markedly different
from Fig.\ 4.  The reason is that $\rho_1(1)\rho_1(2)$ depends
quadratically on $\xi f_i (k)$, whereas $\rho_2 (1, 2)$ depends
on it linearly.  The probability $\xi f_i (k)$ for a hard parton to
be produced at $k$ in HIC is small in central collisions, and is
even smaller in peripheral collisions (although $\xi$ is larger)
because $f_i(k)$ scales with $\left<N_{\rm coll} \right>$.  Thus the
probability of producing two partons in a jet is much higher than
the product of the probabilities of producing the two partons
separately.  The latter does not require the two partons to be
produced in two jets in the same event.  For experiments at RHIC
we may assume that events with two jets in the same hemisphere
never occur, since they are so rare.  Another way of interpreting
$r_2(1,2)$ is to write it as
\begin{eqnarray}
r_2(1,2) = \rho _c (2|1)/\rho_1(2) \quad ,
\label{20}
\end{eqnarray}
where $ \rho _c (2|1)$ is the conditional probability of finding a
parton at $q_2$ given that a parton is at $q_1$
\begin{eqnarray}
 \rho _c (2|1)= {\rho_2 (1, 2) \over \rho_1 (1)} \quad .
\label{21}
\end{eqnarray}
Thus $\rho _c (2|1)$ is like the distribution of a particle
associated with a trigger at  $q_1$.  The largeness of the ratio in
Eq.\ (\ref{20}) means that it is far more likely for a shower
parton to be found at $q_2$ in association with another parton
already found at $q_1$ than to find a parton at $q_2$
independently.

The two surfaces in Fig.\ 5 have essentially the same shape.  The
peripheral case is higher because the corresponding  $\xi f_i
(k)$ is smaller.  In Fig.\ 4 the hard parton momentum $k$ is
held fixed and the probability of having such a parton is 1, while
in Fig.\ 5 $k$ is integrated up to 40 GeV/c so that at higher
$q_1$ and $q_2$ there is no kinematical cut-off, albeit the
probability $f_i(k)$ of producing a hard parton at higher $k$ is
greatly suppressed, which only makes $r_2 (1,2)$ higher.

The study in this section may seem academic, since shower
partons are not directly observable.  However, those shower
partons are the main constituents that contribute to the
formation of pions that we shall treat in the following section. 
Their properties are therefore important to help us understand
the features that we shall find about dihardron correlations. It should be 
emphasized that no free parameters have been used in this section, or in the section to follow, apart from $C, T, \xi$ that have been determined in Sec.\ 2 for the single-particle distributions.

\section{Correlation of Pions in Jets}

For the production of two pions it is necessary to consider the
recombination of four partons.  The corresponding equation for
the two-pion distribution is \cite{hy3}
\begin{eqnarray} 
{dN_{\pi_1 \pi_2}  \over  p_1 p_2 dp_1d p_2} =  {1 \over
p^2_1 p^2_2}
\int\left(\prod^4_{i=1}{dq_i  \over  q_i}\right) F_4
(q_1, q_2,q_3, q_4) R_{\pi_1}(q_1, q_3, p_1)
R_{\pi_2}(q_2, q_4, p_2) .
\label{22}
\end{eqnarray}
The partons with momenta $q_1$ and $q_3$ that form $\pi_1$
can be either thermal or shower, as in Eq.\ (\ref{3}).  Similarly, the same
is true with $q_2$ and $q_4$.  Thus $F_4$ has the form
\begin{eqnarray}
F_4 = ({\cal TT} + {\cal ST} + {\cal SS}) _{13} ({\cal TT} + {\cal
ST} + {\cal SS}) _{24} \quad .
\label{23}
\end{eqnarray}
Among the various possible combinations of terms, it is clear that
$({\cal TT})_{13} ({\cal TT})_{24}$ has nothing to do with hard
scattering and contributes to the background.  
For that reason the terms involving $(\cal TT)$ are not included in \cite{hy3}, where the background-subtracted distributions of the associated particles are considered. Here we study $C_2(1,2)$, which has its own definition of subtraction, so we start out being general.
 Despite the deceptive appearance of  Eq.\
(\ref{23}), $F_4$ is not factorizable.  If it were, its substitution
into Eq.\ (\ref{22}) would render the factorizable form $\rho_2
(1,2) = \rho_1 (1) \rho_2 (2)$, implying no correlation.  The
nonfactorizability of $F_4$ is rooted in the correlations among
the various shower partons.  The correlation between two shower
partons in $({\cal SS}) _{13}$ or $({\cal SS}) _{24}$ has already
been considered in connection with Eq.\ (\ref{19}), which is
shown explicitly in Eqs.\ (\ref{6}) and (\ref{7}).  However, those
are not the correlations that render $F_4$ nonfactorizable.  It is
the correlation between the partons in $(1, 3)$ with the partons
in $(2, 4)$ that gives rise to nonvanishing $C_2 (1,2)$.  More
specifically, because all shower partons in Eq.\ (\ref{23})
originate from the same hard parton at $k$, there is only one   
$\xi f_i (k)$ factor for all ${\cal SS} \cdots$ terms.  Thus, for
example, in $({\cal ST})_{13}({\cal ST})_{24}$ the $({\cal
SS})(q_1,q_2)$ term is exactly as expressed in Eq.\ (\ref{6}).  In 
$({\cal ST})_{13}({\cal SS})_{24}$ we can append
$R_{\pi_2}(q_2,q_4,p_2)$ and make use of Eq.\ (\ref{8}) to
convert $({\cal SS})_{24}R_{\pi_2}$ to the FF $D^{\pi_2}$, so that the
$({\cal ST})({\cal SS})$ contributions to $\rho_2 (1,2)$ is
explicitly 
\begin{eqnarray}
{ d N^{stss}_{\pi_1\pi_2}   \over p_1p_2 dp_1 dp_2} = &&{\xi  \over
p^3_1p_2 } \sum_{i,i',i^{\prime\prime}} \int dk k f_i (k) \int dq_1 {\cal
T}(p_1-q_1)\nonumber\\
&&\cdot \sum_j{1 \over 2} \left[S^j_{i'} \left({ q_1 \over  k} \right){ 1
\over k-q_1 } D^{\pi_2}_{i^{\prime\prime}} \left({ p_2 \over  k-q_1}
\right)+  S^j_{i'} \left({ q_1 \over  k-p_2} \right){ 1 \over
k} D^{\pi_2}_{i^{\prime\prime}} \left({ p_2 \over  k} \right)
\right]
\label{24}
\end{eqnarray}
which is not factorizable.  For definiteness we shall hereafter
consider $\pi_1$ and $\pi_2$ to be both $\pi^+$.  Then the
sum in $j$ is over $u$ and $\bar{d}$.   If $i = g$, then
$i^{\prime},i^{\prime\prime}= g$; if $i = V$ ($u$ or $\bar{d}$), then $(i^{\prime},i^{\prime\prime})
 = (V,S)$ or $(S,V)$ where $S$ denotes sea; if $i = S$, then $i^{\prime},i^{\prime\prime} = S$ also.                               

For completeness, let us write out the two other combinations of
terms.  For $({\cal ST})_{13}({\cal ST})_{24}$ we have
\begin{eqnarray}
{ d N^{stst}_{\pi_1\pi_2}   \over p_1p_2 dp_1 dp_2} = &&{\xi  \over
p^3_1p^3_2 } \sum_{i,i',i''} \int dk k f_i (k) \int dq_1 dq_2 {\cal
T}(p_1-q_1){\cal T}(p_2-q_2)\nonumber\\
&& \sum_{jj^{\prime}} \left\{S^j_{i'} \left({ q_1 \over  k}
\right),  S^{j^{\prime}}_{i''} \left({ q_2 \over  k-q_1}
\right)\right\} \quad ,
\label{25}
\end{eqnarray}
where $j$ and
$j^{\prime}$ can be either $u$ or $\bar{d}$.  For $({\cal
SS})_{13}({\cal SS})_{24}$ we have
\begin{eqnarray}
{ d N^{ssss}_{\pi_1\pi_2}   \over p_1p_2 dp_1 dp_2} = &&{\xi  \over
p_1p_2 } \sum_{i,i',i''} \int dk k f_i (k)\left\{ {1 \over k} D^{\pi_1}_{i'} \left({
p_1 \over  k} \right), 
{1 \over k-p_1} D^{\pi_2}_{i''} \left({
p_2 \over  k-p_1} \right)\right\} \quad .
\label{26}
\end{eqnarray}
In all three equations above $i,i'$ and $i''$ follow the
same rule stated at the end of the previous paragraph.  For
simplicity let us use the symbolic form for these three equations
as follows:  $[({\cal ST}){\cal D}]$, $[({\cal ST})({\cal ST})]$ and
$[{\cal D}{\cal D}]$, respectively.  The parentheses imply the
contents are to recombine; the square brackets signify that the shower partons inside all arise from one hard-parton jet.

 It is appropriate at this point to compare our  description of two-particle distributions with the correlation study of Fries {\it et al} (FBM) \cite{fbm}. Since both approaches are based on the recombination model \cite{hy,hy6,fmn}, there are many similarities, particularly in the separation into various components. Apart from the difference in focus where we consider correlation in $p_T$, while FBM studies correlation in $\phi$, the main difference in physics is that we break the hard parton into shower partons, whereas FBM does not. If our $\cal TS$ recombination is related qualitatively to their soft-hard (SH) recombination, then the classification of components is essentially the same in the two approaches. Quantitatively, $\cal TS$ can be significantly different from SH. Moreover, the correlation at the parton level is for us among the shower partons, while for FBM among the soft partons.

In $\rho_2 (1,2)$ the $({\cal TT})({\cal TT})$ term is
factorizable, and is cancelled by the thermal terms in $\rho_1
(1) \rho_1 (2)$, where $\rho_1 (1)$ is given by Eq.\ (\ref{10}). 
That corresponds to background subtraction.  Other double
${\cal TT}$ terms like $({\cal TT})({\cal ST})$ also get cancelled.  The
surviving terms in $C_2 (1,2)$ all involve shower partons in each
pion.  We group them in four clusters of terms
\begin{eqnarray}
C_2 (1,2) = C^{stst}_2 (1,2) + C^{stss}_2 (1,2) + C^{ssst}_2 (1,2)+
C^{ssss}_2 (1,2) \quad ,  
\label{27}
\end{eqnarray}
where
\begin{eqnarray}
C^{stst}_2 (1,2) = \left[({\cal ST})({\cal ST}) \right]_{12} - ({\cal
ST})_1* ({\cal ST})_2
\label{28}
\end{eqnarray}
\begin{eqnarray}
C^{stss}_2 (1,2)= \left[({\cal ST})({\cal D}) \right]_{12} - ({\cal
ST})_1* ({\cal D})_2
\label{29}
\end{eqnarray}
\begin{eqnarray}
C^{ssst}_2 (1,2) = \left[{\cal D}({\cal ST}) \right]_{12} - ({\cal
D})_1 *({\cal ST})_2 
\label{30}
\end{eqnarray}
\begin{eqnarray}
C^{ssss}_2 (1,2) = \left[{\cal D}{\cal D} \right]_{12} - ({\cal
D})_1 *({\cal D})_2  \quad .
\label{31}
\end{eqnarray}
The products involving $*$ signify terms from $\rho _1(1)\rho
_1(2)$, so their shower partons have their own hard partons. 
More precisely, $({\cal ST})_1* ({\cal ST})_2$, for example, has
the explicit expression
\begin{eqnarray}
({\cal ST})_1* ({\cal ST})_2 = \left({\xi  \over
p^3_1} \sum_{ij} \int dk k f_i (k)\int  dq_1
S^j_i \left({ q_1 \over  k} \right) {\cal
T}(p_1-q_1)\right)\nonumber\\
\cdot \left({\xi  \over
p^3_2} \sum_{ij} \int dk^{\prime} k^{\prime} f_i
(k^{\prime})\int  dq_2 S^j_i \left({ q_2
\over  k^{\prime}}
\right) {\cal T}(p_2-q_2) \right) \quad .
\label{32}
\end{eqnarray}
Clearly, this is very different from $\left[({\cal ST})({\cal ST})
\right]_{12}$, written out in full in Eq.\ (\ref{25}).  Thus
$C^{stst}_2 (1,2)$ does not vanish, in general, and likewise for
the other terms in Eq.\ (\ref{27}).

 Regarding Eqs.\ (\ref{26}) and (\ref{31}) that involve FFs only, it is relevant to compare them to the medium modified dihadron FF studied in \cite{mww}. In our formulation of hadron production in HIC the medium effect is taken into account by considering $(\cal TT)$ and $(\cal TS)$ recombination, leaving the $(\cal SS)$ recombination to be unaffected by the medium except for the hard parton distribution $f_i(k)$. Thus in Eqs.\ (\ref{9}) and (\ref{26}) the FF in vacuum is used. The contribution of $C_2^{ssss}(1,2)$ to the observed correlation cannot be measured separately, except in $pp$ collisions or in very peripheral $AA$ collisions where the medium effect is negligible.  In the fragmentation model modification of the FF is the only way to take into account of the medium effect, whereas in the recombination model fragmentation in vacuum is unimportant compared to thermal-shower recombination. For these reasons  it is inappropriate to compared Eq.\ (\ref{26}) for central nuclear collisions with the medium modified dihadron fragmentation process  considered in \cite{mww} without including Eqs.\ (\ref{24}) and (\ref{25}) also.

We have calculated all the quantities in Eqs.\ (\ref{27}) -
(\ref{31}).  The results cannot easily be displayed because each
term decreases rapidly with $p_1$ and $p_2$ and the various
clusters of terms, $C^{\alpha}_2$, $\{\alpha = stst, \dots \}$, can
be positive and negative, so that neither linear nor log plots are
appropriate.  To give a sense of the complication, we show in
Fig.\ 6(a) the result for $\pi^+\pi^+$ correlation in central Au+Au collisions for the restricted case of $p_1 = p_2 = p$ in the
range 1.5-10 GeV/c.  At $p \approx 2$ GeV/c, all $C^{\alpha}_2$
become negative, resulting in a dip in $C_2$.  There is no such fine
structure for peripheral collisions (80-92\%), so a semi-log plot can
exhibit the behaviors of $C^{\alpha}_2$ in the entire range up to $p =
10$ GeV/c, as shown in Fig.\ 6(b).  Note that for 0-10\% centrality the
terms involving $st$ component is dominant, while for 80-92\%
centrality the $\alpha = ssss$ component is dominant.  These features are in
accord with the properties of single-particle distributions shown in
Fig.\ 3, where thermal-shower and shower-shower components,
respectively, are dominant in the two cases.

When $p_1$ and $p_2$ are different, the two surfaces of $C_2
(1,2)$ for the two centralities are shown in Fig.\ 7 in the 2D ranges
where they are positive. The region where $C_2(1,2)$ is negative cannot be shown in the semi-log plot.  For what can be shown they behave rather
similarly.  They both are suppressed by about 10 orders of magnitude
going from 1 to 10 GeV/c.  A large part of the rapid decrease is
due to the power-law suppression that is present even in the single-particle distributions over the
same $p_T$ range, as can be seen in Fig.\ 1, though not more than 8
orders of magnitude.  Thus the correlation function $C_2
(1,2)$ is not the best quantity to exhibit the properties of
correlation.

To de-emphasize the power-law suppression of the hard parton
distribution at high $k$ that is contained in $\rho_1(1)$ and
$\rho_1(2)$, we consider the normalized correlation $K_2
(1,2)$ defined in Eq.\ (\ref{13}).  When we set $p_1 = p_2 = p$,
the dependence of $K_2(p)$ on $p$ is shown in Fig.\ 8 for (a)
0-10\% and (b) 80-92\% centrality.  In (a)  a dip occurs
at low $p$, the minimum being at $p = 3$ GeV/c  with
$K_2 < 0$;  in (b) there is no dip.  The rapid rise at larger $p$ is
introduced into $K_2 (1,2)$ by the division of $C_2 (1,2)$ by
$\rho_1(1)\rho_1(2)$, which is strongly damped at high $p_1$
and $p_2$.  It is the result of an over-correction.  The fact that
$K_2 (80-92\%) \gg K_2 ({\rm 0-10}\%)$ at all $p$ in Fig.\ 8  is due
to the property of $\xi f_i(k)$ being much smaller in peripheral
collisions compared to central collisions.  Since the numerator of
$K_2(1,2)$ depends linearly on $\xi f_i(k)$, but the
denominator depends quadratically on it, the peripheral
$K_2(1,2)$ is therefore much larger.

What we see in Fig.\ 8 (a) and (b) about the general features of $K_2(1,2)$
for the two centralities is very similar to what is shown in Fig.\ 5 for
$r_2 (1,2)$ for shower partons, recalling that $K_2(1,2) = r_2
(1,2)-1$, as stated in Eq.\ (\ref{13}).  Although $r_2(1,2)$ cannot become negative, it becomes large  in Fig.\ 5 at high $q_1$ and $q_2$ for both centralities.  Of course,
hadron correlation is far more complicated than parton
correlation.  But the common feature that they both become
large at large parton and hadron momenta is due to the
commonality in the definition of $r_2 (1,2)$ and $K_2(1,2)$,
which have $\rho_1(1)\rho_1(2)$ in the denominators.

It is now clear that we need something that is between $C_2
(1,2)$ and $K_2(1,2)$ so as to have linear dependence on $\xi
f_i(k)$ in both the numerator and the denominator.  That
quantity is the ratio $G_2(1,2)$, defined in Eq.\ (\ref{15}). 
Figure 8 (c) and (d) shows the dependence of $G_2(1,2)$ on
$p_1=p_2=p$ for (c) 0-10\% and (d) 80-92\% centrality.  Now,
$G_2(p)$ is small at large $p$ in both cases, and the properties of both can be revealed in linear plots.  For 0-10\% centrality there is still a dip
at $p = 2$ GeV/c with $G_2(p) < 0$ in its vicinity, as it should, but for 
80-92\% centrality $G_2(p)$ decreases monotonically with
increasing $p$.  The normalizations of  $G_2(p)$ for the two
cases are now nearly the same.  That can be seen more easily in
semi-log plots.  In Fig.\ 9 we show all three correlation function,
$C_2$, $K_2$ and $G_2$, for the two centralities.  The gap in
Fig.\ 9(a) is due to the functions being negative in that range. 
Evidently, we have $C_2 (0-10\%) \gg C_2 (80-92\%) $, $K_2
(0-10\%) \ll K_2 (80-92\%) $, but $G_2 (0-10\%) \approx G_2
(80-92\%) $ in regions where they can be compared.  Thus the
function $G_2 (1,2)$ is best for the study of the nature of
correlation.

In Fig.\ 10(a) we show $G_2 (1,2)$ in 2D distributions for both
centralities in the same linear space.  We now can see clearly
their differences, knowing that asymptotically they behave
similarly without structure.  In the ranges of $p_1$ and $p_2$
plotted $G_2 (1,2)$ shows negative correlation for 0-10\%, but
positive correlations for 80-92\% centrality.  That difference can
be succinctly exhibited in the ratio
\begin{eqnarray}
R^{G_2}_{CP}(1,2) = { G^{(0-10\%)}_2 (1,2)\over
G^{(80-92\%)}_2 (1,2)} ,
\label{33}
\end{eqnarray}
which is shown in Fig.\ 10(b).  Since peripheral Au+Au
collisions at 80-92\% centrality is very close to $pp$ collisions,
we may regard Fig.\ 10(b) as the clearest revelation of the
difference in dihadron correlation between central Au+Au and
$pp$ collisions.

The origin of the dip in Fig.\ 10(b) can be found in Fig.\ 6
where $C_2$ is negative at $p_1 = p_2 = p =2$ GeV/c for
0-10\% centrality, but not for 80-92\% centrality.  
 Let us now describe the reasons why $C_2^\alpha$ are negative in Fig.\ 6(a), but not in Fig.\ 6(b).  Many factors are involved with competing strengths that can be compared only by concrete calculation, but qualitative discussion can provide some insight. Consider Eq.\ (\ref{28}) whose two terms are shown explicitly in Eqs.\ (\ref{25}) and (\ref{32}). The former has one factor of $\xi f_i(k)$ and two SPDs sharing the same $k$, while the latter has two independent factors of $\xi f_i(k)$, each having only one SPD. In the peripheral case $\xi f_i(k)$ is very small because $f_i(k)$ is scaled down by $\left<N_{\rm coll}\right>$, so $({\cal ST})_1*({\cal ST})_2$ is always small compared to $[({\cal ST})({\cal ST})]_{12}$ at all $q_1$ and $q_2$. Hence, $C_2^{stst}(1,2)$ is always positive; the same argument applies to all other $C_2^{\alpha}(1,2)$. In the case of central collisions $\xi f_i(k)$ becomes large enough to allow other factors to also play some role. When $p_1$ and $p_2$ are on the lower end of the intermediate range, e.g. $\le 4$ GeV/c, $q_1$ and $q_2$ must be even less, but not too small because ${\cal T}(p_1-q_1)$ and ${\cal T}(p_2-q_2)$ would be exponentially damped at large $p_i-q_i$. In other words, if the momentum fraction $x_i$ of the shower partons are small, there are not many thermal partons to make up $p_1$ and $p_2$; on the other hand, if $x_i$ are large, the SPDs are suppressed. The hard parton $k$ cannot be large to relieve the latter constraint without suffering the suppression of $f_i(k)$. This situation of multiple constraints is easier for $({\cal ST})_1*({\cal ST})_2$ to meet since there is only one shower parton per jet in Eq.\ (\ref{32}), but is much more difficult for $[({\cal ST})({\cal ST})]_{12}$ where the jet momentum must be shared by two shower partons. Thus $[({\cal ST})({\cal ST})]_{12}$ is forced to be smaller than $({\cal ST})_1*({\cal ST})_2$ in central collisions where $\xi f_i(k)$ is not infinitesimal and thermal partons are important in the intermediate $p_T$ region.

 In the case of $C_2^{ssss}(1,2)$ there is no $\cal T$, but the demand for additional shower partons plays the same role. It is clear from Eq.\ (\ref{31}) that  double fragmentation causes $[{\cal DD}]_{12}$ to be more suppressed than two single fragmentation $({\cal D})_1*({\cal D})_2$. Thus 
  $C_2^{ssss}(1,2)$ becomes negative so long as $p_1$ and $p_2$ are not high enough to force the large $k$ to dampen double $\xi f_i(k)$ in $({\cal D})_1*({\cal D})_2$ more severely than the single $\xi f_i(k)$ in $[{\cal DD}]_{12}$. The other $C_2^\alpha$ terms behave similarly for the same reason, resulting in the sum $C_2(1,2)$ to be negative when $p_1$ and $p_2$ are both less than around 3 GeV/c.  The negativity of $C_2(1,2)$ should not be surprising, since the ratio $r_2(1,2)$ for the shower partons in a jet of fixed $k$ is less than 1, as shown in Fig.\ 4. The corresponding $C_2(1,2)$ is negative for all $x_1$ and $x_2$.   In the hadronic problem there are in some terms even more shower partons involved, all competing for the limited momentum that the hard parton provides. Fig.\ 10 exhibits the intricacies of the hadronic correlation even when the basic correlations among the shower partons are kinematical in nature.
 
\section{Comparison with Experimental Data}

In the foregoing we have studied the correlation of two $\pi^+$
without using any adjustable parameters. What we have obtained
unfortunately cannot be compared directly with current data, because
most experiments either do not yet have particle identification, or
put emphasis on jets with trigger particles and study the
dependencies on $\Delta\phi$ and $\Delta\eta$ of associated
particles. There are, however, two areas  in which our theoretical
calculations come close to the experimental data and are of interest
for us to examine them in more detail here.

The first area concerns the experiment in \cite{ja2}, in which data
are given for the $p_T$ distribution of charged particles associated
with trigger in the range $4<p_T<6$ GeV/c. We can come close to what
is measured by calculating
\begin{eqnarray}
{dN_{\pi^+}^{tr-bg}\over
p_2dp_2}={\int_4^6dp_1p_1{dN_{\pi^+\pi^+}^{tr-bg}\over
p_1p_2dp_1dp_2}/ \int_4^6dp_1p_1{dN_{\pi^+}\over
p_1dp_1}},   \label{35}
\end{eqnarray}
where $p_1$ is regarded as the momentum of the trigger particle, and
$p_2$ that of the associated particle. The superscript, $tr-bg$,
denotes the sample of triggered events  with background
subtracted. For the integrand in the numerator we use Eq.\
(\ref{22}), but with $F_4$ not given by Eq.\ (\ref{23}). Background
subtraction corresponds in our case to leaving out in Eq.\
(\ref{23}) the terms that are factorizable. We are uncertain whether
that correspondence is exact, since the experimental procedure for
the determination of the background is not unique and involves steps
that seem hard to duplicate in theoretical calculation. Our
procedure of retaining only the non-factorizable component is the
most sensible way to obtain the associated particle distribution,
so we use
\begin{eqnarray}
F_4^{tr-bg}={\cal (ST+SS)}_{13}{\cal(ST+SS)}_{24},  \label{36}
\end{eqnarray}
which is not factorizable because of the correlation that
exists between the shower parton at $q_1$ and the shower parton at
$q_2$, as we have studied in Sec.\ 3.

We note that Eq.\ (\ref{35}) is a ratio of the integrals over the
trigger momentum in contrast to the integral of the ratio that is
calculated in \cite{ hy3}. Eq.\ (\ref{35}) corresponds to the way
that the data are presented in \cite{ja2}.

There is one piece of detail about the calculation using Eqs.\
(\ref{22}) and (\ref{36}) that we now discuss. Among the SPDs
determined in \cite{hy4} the gluon initiated quark distribution has a
parametrization that leads to divergence, $G(x)\propto x^{-a}$, as
the momentum fraction $x\rightarrow 0$. It is unphysical in that
limit, although the parametrization renders a good fit of the
fragmentation function away from that limit. In application to the
calculation of $dN_{\pi}/pdp$, as in Eq.\ (\ref{10}), the integral
over $q_1$ is convergent, and at small $p$ the $\cal ST$ component
is dominated by the $\cal TT$ component. But when the background is
subtracted with the $\cal (TT)(TT)$ component removed from
$dN_{\pi\pi}/p_1p_2dp_1dp_2$, the effect of large $G(x)$ at very
small $x$ is exposed. Since that effect is unphysical and since we
do not claim reliability of our formalism at very small $p_T$, we
introduce an {\it ad hoc} factor
\begin{eqnarray}
s(q)=1-e^{-q/q_0}   \label{37}
\end{eqnarray}
to suppress the distribution at small $q$ for all gluon initiated
shower partons. With $q_0$ set at 1 GeV/c this procedure does not affect the results of our
calculation for $p_T>1.5$ GeV/c, but does introduce a round-off at
lower $p_T$ in the associated particle distribution. Figures in Sec.\ 4 
all have that factor $s(q)$ included for gluon initiated showers.

Substituting Eq.\ (\ref{36}) into Eq.\ (\ref{22}) and then into Eq.\
(\ref{35}), we obtain the results shown in Fig.\ 11 for (a) 0-10\%
and (b) 80-92\% centralities. The calculated curves are for $\pi^+$
trigger and $\pi^+$ associated particle, while the data points from
\cite{ja2} are for all charged particles with 0-5\% centrality in (a)
and for $pp$ collision in (b). The shapes of the $p_T$ dependence
agree very well; the normalizations differ by  a factor that can 
be attributed to the data being for all charged particles. Thus our formalism
describes the correlation between particles in jets quite satisfactorily when
compared to data.

Another area of possible comparison between theory and experiment is
in correlation at low $p_T$. Data on a quantity denoted by
$\Delta\rho/\sqrt \rho$ in Ref.\ \cite{jp} for $pp$ collision show a
bulge plotted in transverse rapidity, defined by
\begin{eqnarray}
y_t=\ln [(m_T+p_T)/m_{\pi}].     \label{38}
\end{eqnarray}
Since the quantity $\Delta\rho/\sqrt \rho$ is rather similar to our
$G_2(1,2)$, it is of interest to display the latter in terms of
$y_t$, which has the virtue of expanding the low $p_T$ scale.  Since our calculations are
unreliable for $p_T<1$ GeV/c, we consider only the region $y_t>2.5$.
Instead of Eq.\ (\ref{12a}), we now define
\begin{eqnarray}
\rho_1(1)={dN_{\pi_1}\over dy_1},  \hskip1cm
\rho_2(1,2)={dN_{\pi_1\pi_2}\over dy_1dy_2},   \label{39}
\end{eqnarray}
where $dy_i=dp_i/m_T$. $C_2(1,2)$ and $G_2(1,2)$ remain to be
defined by Eqs.\ (\ref{12}) and (\ref{15}), respectively, but in
terms of $y_1$ and $y_2$ they are changed in accordance to Eq.\
(\ref{39}). 

In Fig.\ 12 we show $G_2(1,2)$ as functions of $y_1$ and
$y_2$ for (a) 80-92\% and (b) 0-10\% centralities. They are
reproductions of the two surfaces in Fig.\ 10(a), now in terms of
$y_i$. The upper ones in both are for peripheral collisions, while
the lower ones are for central collisions. The increase at small $y_1$ and $y_2$ in Fig.\ 12(a)
appears to agree with the bulge in the $pp$ collision data in
\cite{jp} for $y_1,y_2>2.5$. For central collision Fig.\ 12(b) has a
peak and a dip, which are predictions that should be checked by the
analysis of the corresponding data. For completeness, we show in Fig.\ 13 the $R_{CP}^{G_2}$ in the $y_1, y_2$ variables, in terms of which the dip becomes the dominant feature, even more prominent than in Fig.\ 10 (b).

\section{Conclusion}

We have calculated the single-particle distribution in
Au+Au collisions at all centralities, and considered various types of correlation functions in jets.  For correlations between shower partons we have shown the drastic difference between the
case when the hard parton momentum $k$ is fixed, as in
$e^+e^-$ annihilation, and the case in heavy-ion  collisions
where $k$ is not fixed.  For hadron correlations we have
considered three types of correlation fucntions, $C_2 (1,2)$,
$K_2 (1,2)$, and $G_2 (1,2)$, and showed that $G_2 (1,2)$ is
best in displaying the structure of the dihadron correlation.

We have found that the ratio of $G_2 (1,2)$ for central collisions to
that for peripheral collisions has a big dip in the $p_1
\approx p_2 \approx 2$ GeV/c region.  That is where the negative correlation
between shower partons in a jet pulls down the hadronic $\rho_2(1,2)$ in central Au+Au collisions relative to $\rho_1(1)\rho_1(2)$. 
However, in peripheral collisions $\rho_1(1)\rho_1(2)$ is so small compared to  $\rho_2(1,2)$ that $C_2(1,2)$ remains positive for all $p_1$ and $p_2$. The calculated
features of $G_2 (1,2)$ can be checked by experiment, since all the
quantities involved are directly measurable.

Although most of the results obtained on the correlation functions
cannot be compared with any existing data, there is no basic hurdle
to overcome for the data analysis to be carried out in ways that can
check our prediction. The distribution of associated particles that
we have calculated for $\pi^+\pi^+$ correlation compares
satisfactorily with the data except for the normalization difference
due to the fact that the data are for all charged particles. Our
result on $G_2$ for peripheral collisions has some resemblance to
the data on $pp$ collisions, which are, however, not presented in a
way that can facilitate quantitative comparison. Clearly, closer
coordination between theoretical and experimental efforts would be
of great value to both.

In this work the basic quantity is the correlation function $C_2
(1,2)$ that is precisely defined.  By subtracting
$\rho_1(1)\rho_1(2)$ from $\rho_2(1, 2)$, not only is the
background subtracted, other factorizable terms are taken out
too, creating the possibility that $C_2 (1,2)$ can become
negative.  That possibility does not show up in the present experimental
way of determining the associated particle distribution
\cite{dm,ja2}, but may be present in other ways of studying
correlations \cite{tt,jp,ja}.  We suggest that future experimental
work be directed toward determining the correlation function 
$G_2 (1,2)$  and the central-to-peripheral ratio
$R^{G_2}_{CP}(1,2)$.  The absence of any hole in the
experimental result on $R^{G_2}_{CP}(1,2)$, as shown in Figs.\
10(b) and 13, would provide valuable information on the nature of parton correlation that we have not incorporated in our formulation of the  problem.

\section*{Acknowledgment}
We are grateful to  Tom Trainor and Fuqiang Wang for
valuable communication on their work and to C.\ B.\ Yang for helpful comments.  This work was
supported, in part,  by the U.\ S.\ Department of Energy under
Grant No. DE-FG03-96ER40972.

\newpage
\begin{center}
\section*{Figure Captions}
\end{center}

\begin{description}
\item
Fig.\ 1. Single-particle distributions of produced $\pi$ in Au+Au
collisions at $\sqrt s=200$ GeV for four centrality cuts. Data are
from \cite{ph}, and the solid lines are the results of calculation
in the recombination model.

\item
Fig.\ 2. The three parameters $C, T$ and $\xi$, determined by fitting
the data at all centralities.

\item
Fig.\ 3. The single-particle distributions for (a) most central and
(b) most peripheral collisions, showing the three components that
contribute to the sums in solid lines.

\item
Fig.\ 4. (Color online) The ratio $r_2(1,2)$ in terms of the  shower parton momentum
fractions
$x_1$ and $x_2$ for a hard gluon momentum fixed at a large values
$k$.

\item
Fig.\ 5. (Color online) The ratio $r_2(1,2)$ in terms of the  shower parton momenta
$q_1$ and $q_2$ in heavy-ion collisions for two extreme centralities.

\item
Fig.\ 6.  The correlation functions $C_2$ for $\pi^+\pi^+$ in HIC,
showing the contributions from the three combinations of $\cal S$
and $\cal T$ partons for (a) central and (b) peripheral collisions
along the diagonal where $p_1=p_2$.

\item
Fig.\ 7. (Color online) The correlation functions $C_2$ for two centralities when
they are positive.

\item
Fig.\ 8.  The normalized correlation functions $K_2$ along the
diagonal where $p_1=p_2$ for (a) central and (b) peripheral
collisions. The partially normalized correlation functions $G_2$  
for (c) central and (d) peripheral collisions.

\item
Fig.\ 9.  A comparison of the three types of correlation functions
$C_2, K_2$ and $G_2$ in semi-log plots, with the gap corresponding
to where the functions are negative.

\item
Fig.\ 10. (Color online)  (a) $G_2(1,2)$ plotted as functions of $p_1$ and $p_2$
for two centralities; (b) the ratio $R_{CP}^{G_2}$ of central to
peripheral $G_2$.

\item
Fig.\ 11.  Associated particle distributions of $\pi^+\pi^+$ for (a)
central (0-10\%) collisions compared to the STAR data \cite{ja2} for
all charged particles at (0-5\%) centrality and (b) peripheral
(80-92\%) collisions compared to the STAR data \cite{ja2} for all
charged particles in $pp$ collisions.

\item
Fig.\ 12.  (Color online)  $G_2(1,2)$ for $\pi^+\pi^+$ plotted in terms of transverse
rapidities for (a) peripheral (80-92\%) and (b) central (0-10\%)
collisions.

\item
Fig.\ 13. (Color online) The ratio $R_{CP}^{G_2}$ in transverse rapidities.

\end{description}

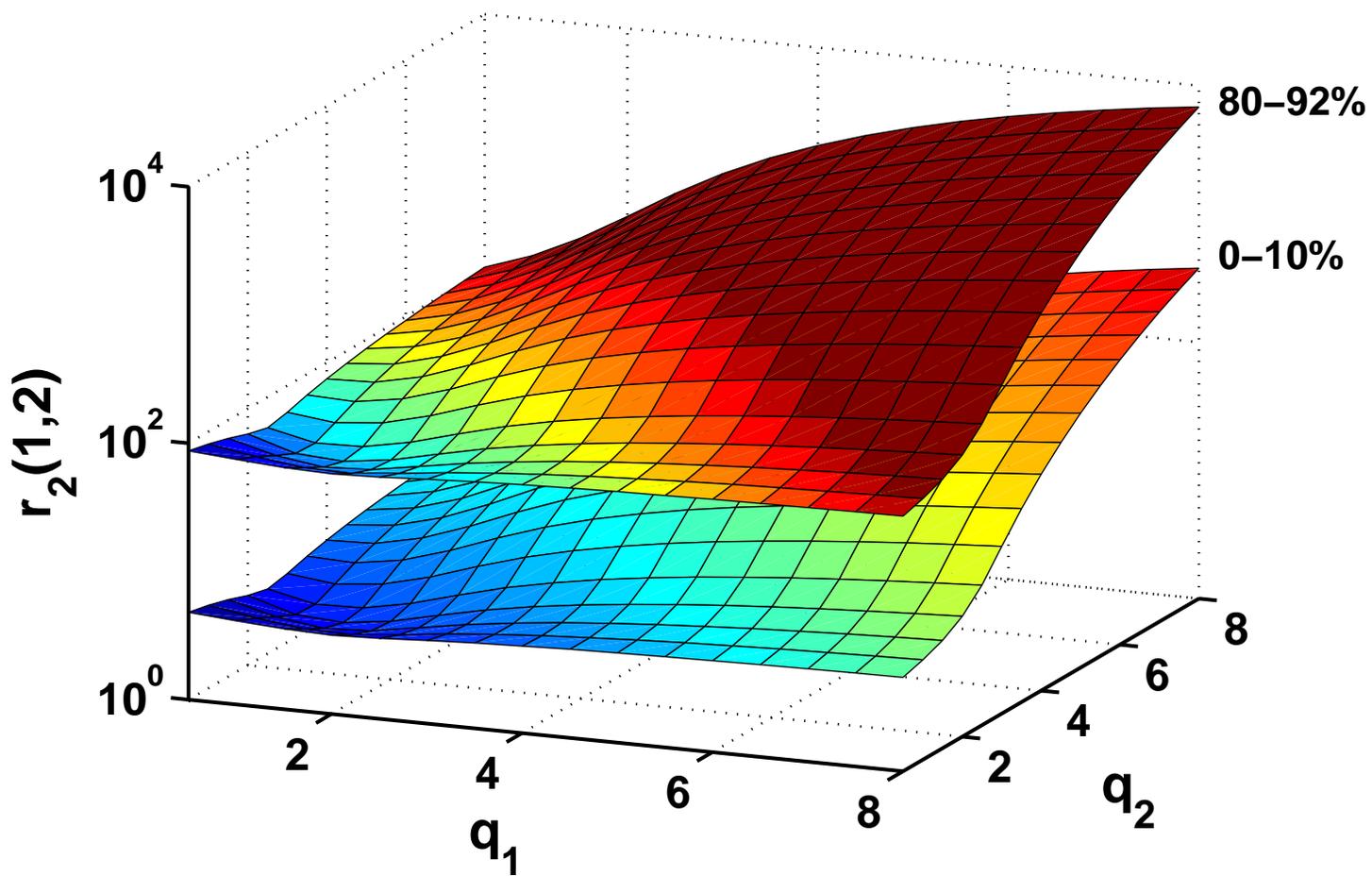
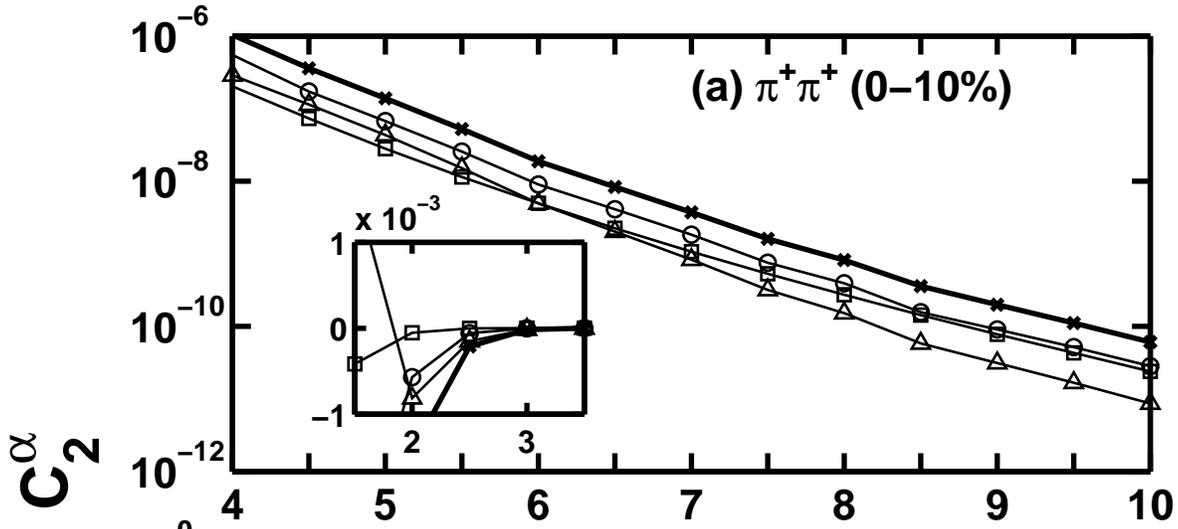
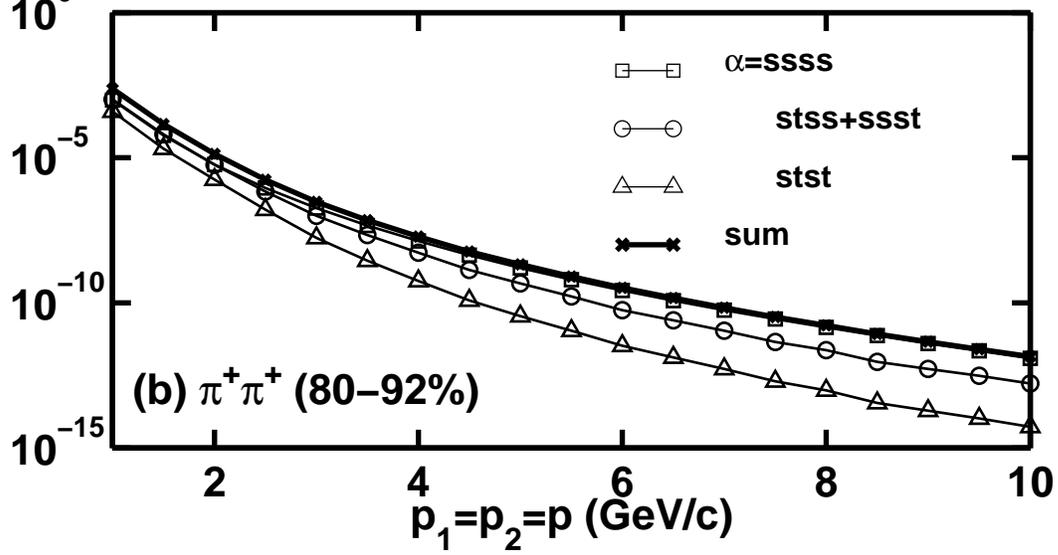
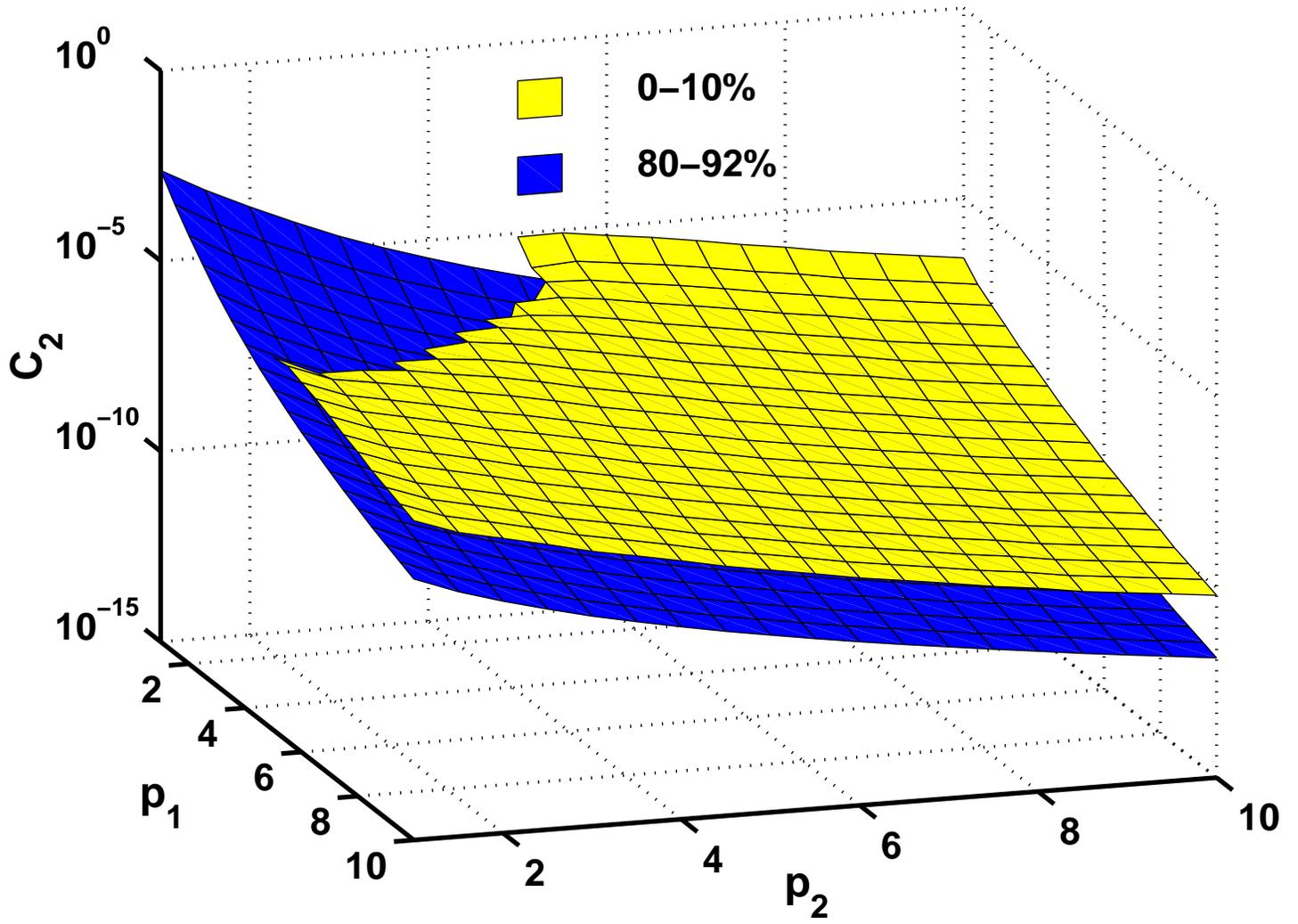
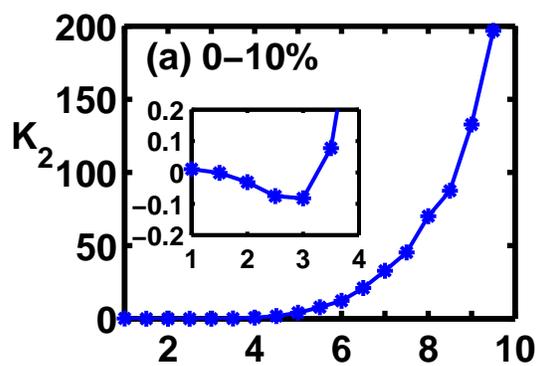
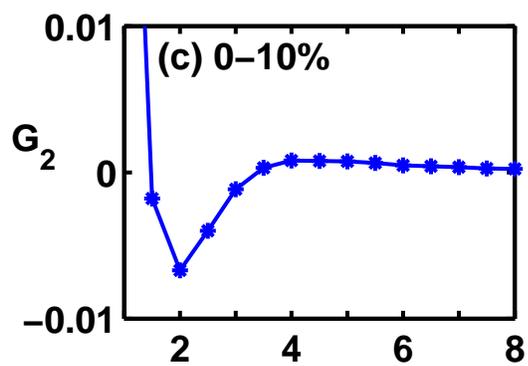
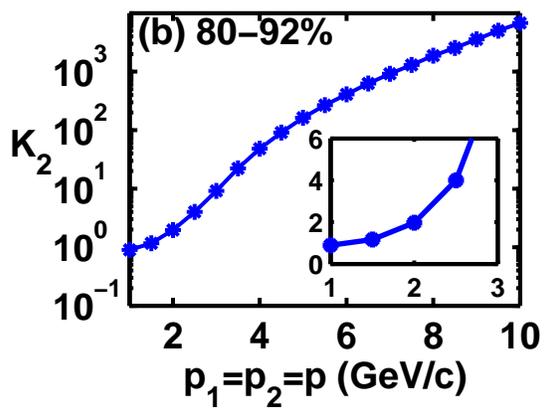
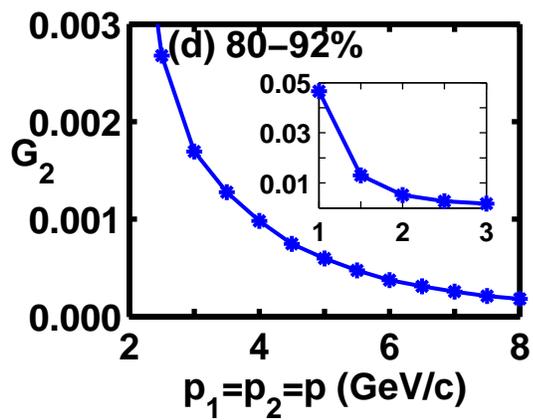
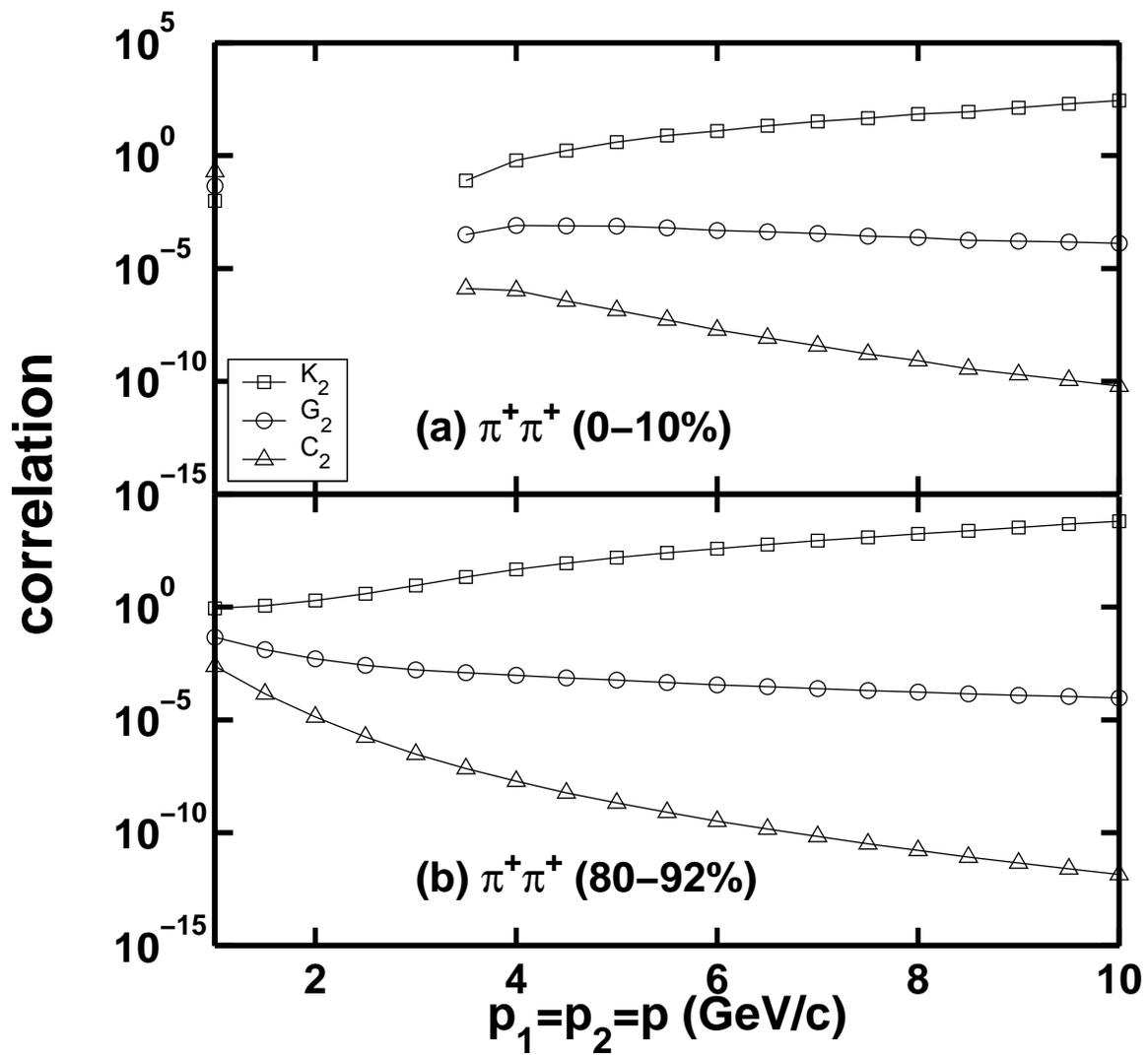
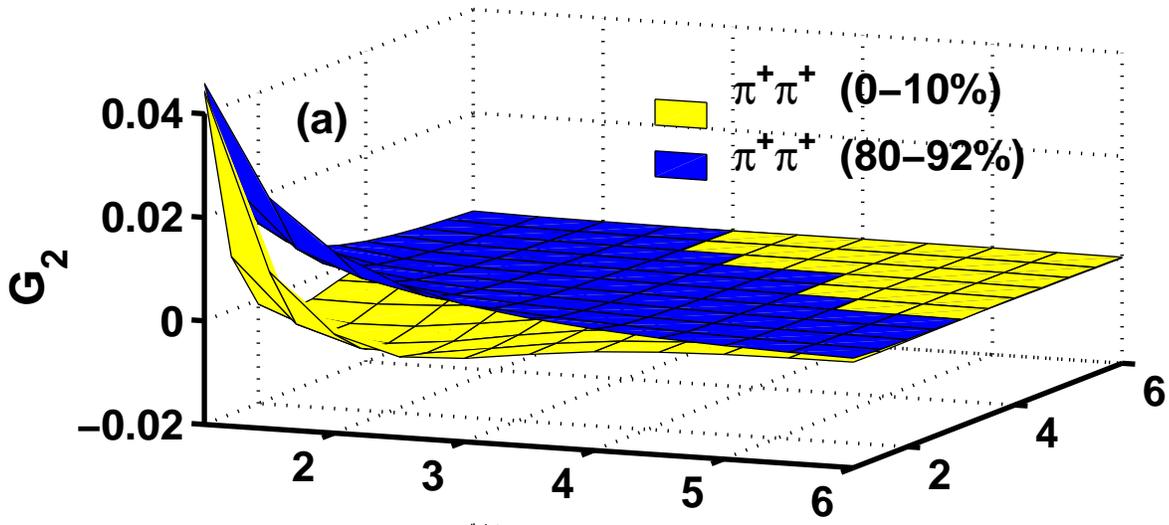
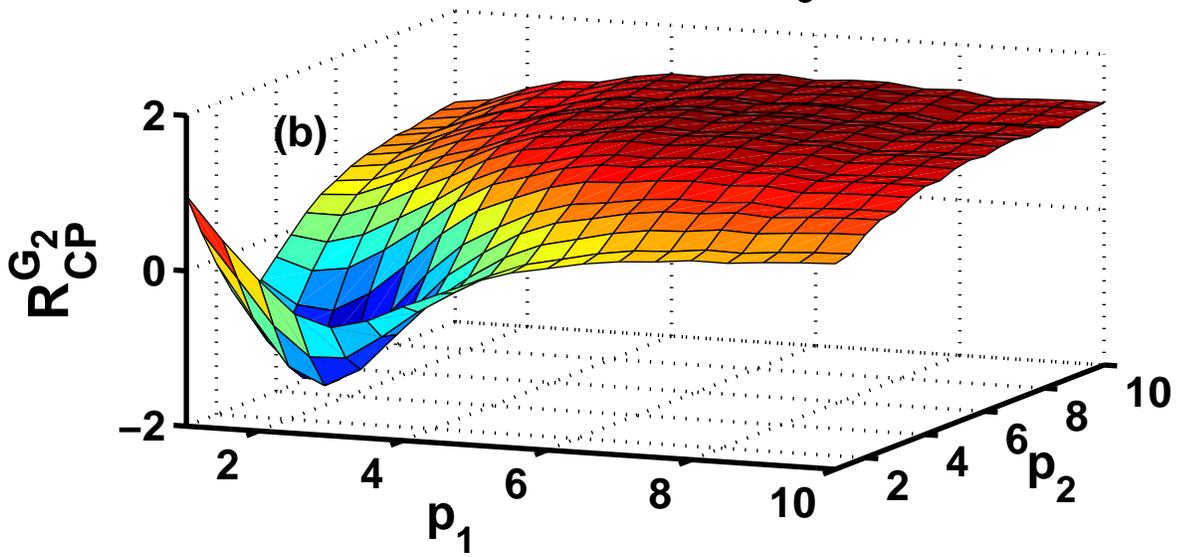
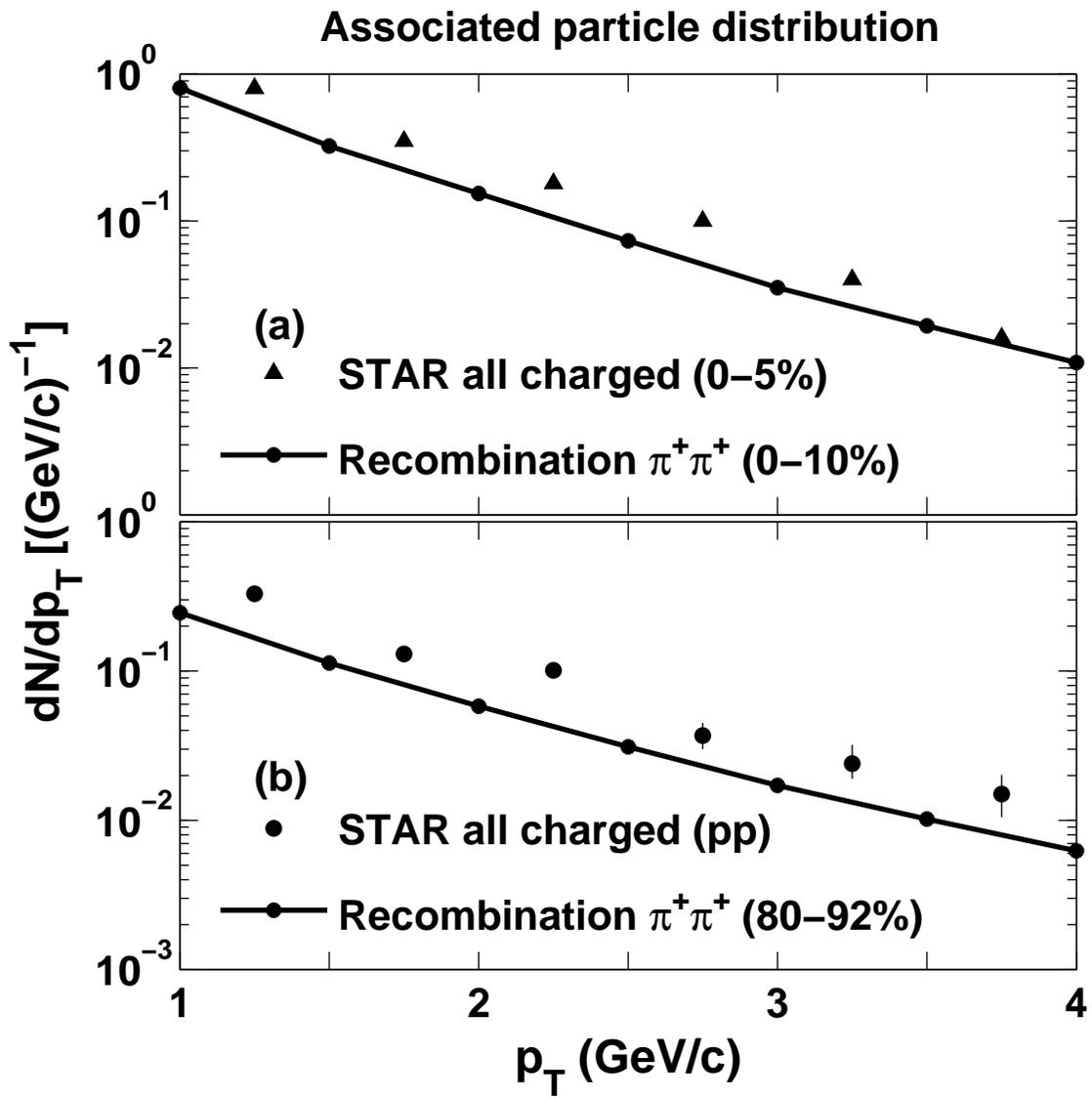
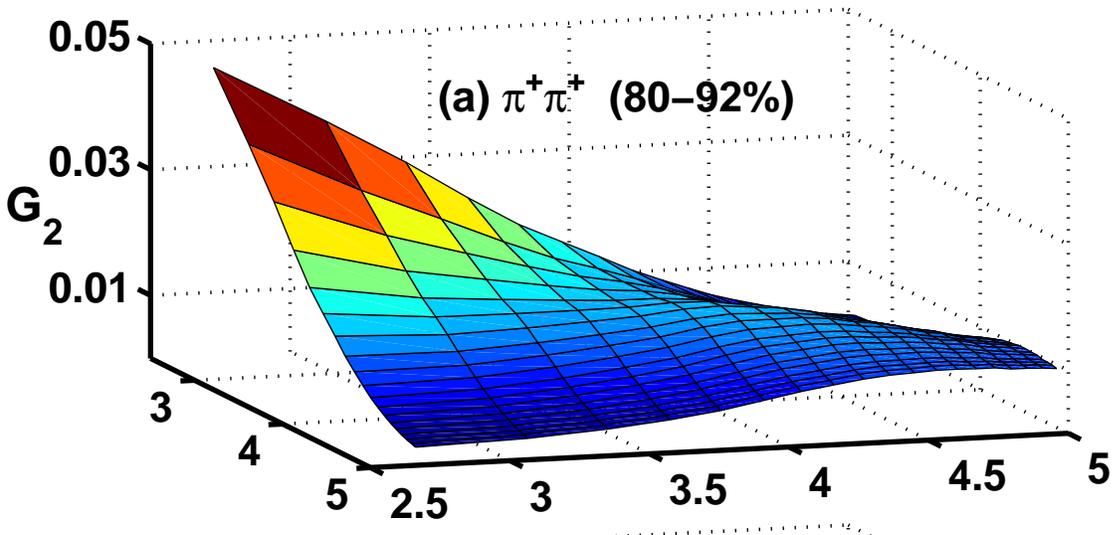
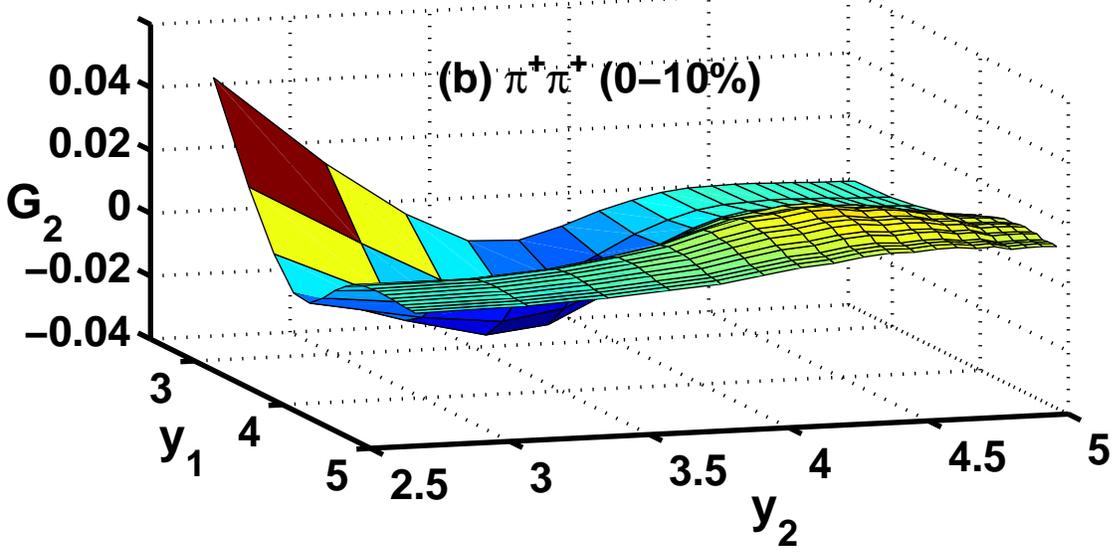
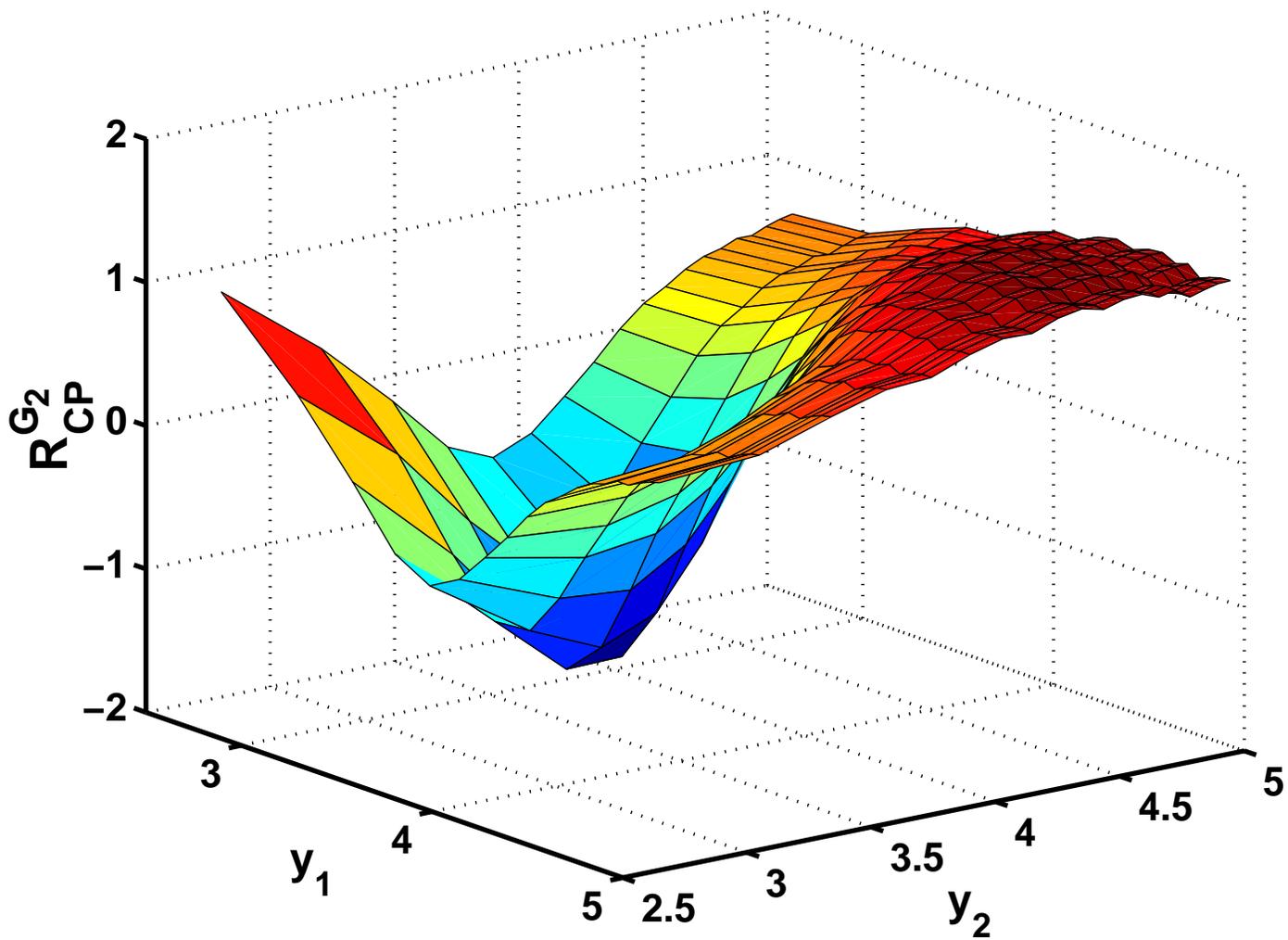
\end{document}